\documentclass[twocolumn,aps,prc,showpacs,superscriptaddress,preprintnumbers,floatfix,nofootinbib]{revtex4}
\usepackage{epsfig,graphics}
\usepackage{graphicx}
\usepackage{dcolumn}
\usepackage{bm}
\usepackage{amsmath}
\usepackage[usenames]{color}
\usepackage{ulem} 

\begin{document}

\title{Transverse momentum and pseudorapidity dependences of $'\bm{Mach-like}'$ correlations for central Au + Au collisions at $\sqrt{s_{NN}}$ = 200 GeV}

\author{ S. Zhang}
\affiliation{Shanghai Institute of Applied Physics, Chinese Academy of Sciences, P.O. Box 800-204, Shanghai
201800, China} \affiliation{Graduate School of the Chinese Academy of Sciences, Beijing 100080, China}
\author{ G. L. Ma}
\affiliation{Shanghai Institute of Applied Physics, Chinese Academy of Sciences, P.O. Box 800-204, Shanghai
201800, China}
\author{ Y. G. Ma}
\thanks{Corresponding author: Email: ygma@sinap.ac.cn}
\affiliation{Shanghai Institute of Applied Physics, Chinese Academy of Sciences, P.O. Box 800-204, Shanghai
201800, China}
\author{ X. Z. Cai}
\affiliation{Shanghai Institute of Applied Physics, Chinese Academy of Sciences, P.O. Box 800-204, Shanghai
201800, China}
\author{ J. H. Chen}
\affiliation{Shanghai Institute of Applied Physics, Chinese Academy of Sciences, P.O. Box 800-204, Shanghai
201800, China} \affiliation{Graduate School of the Chinese Academy of Sciences, Beijing 100080, China}
\author{H. Z. Huang}£¬
\affiliation{ Dept of Physics and Astronomy, University of California at Los Angeles, CA 90095, USA}
\author{ W. Q. Shen}
\affiliation{Shanghai Institute of Applied Physics, Chinese Academy of Sciences, P.O. Box 800-204, Shanghai
201800, China}
\author{ X. H. Shi}
\affiliation{Shanghai Institute of Applied Physics, Chinese Academy of Sciences, P.O. Box 800-204, Shanghai
201800, China} \affiliation{Graduate School of the Chinese Academy of Sciences, Beijing 100080, China}
\author{ F. Jin}
\affiliation{Shanghai Institute of Applied Physics, Chinese Academy
of Sciences, P.O. Box 800-204, Shanghai 201800, China}
\affiliation{Graduate School of the Chinese Academy of Sciences,
Beijing 100080, China}
\author{ J. Tian}
\affiliation{Shanghai Institute of Applied Physics, Chinese Academy
of Sciences, P.O. Box 800-204, Shanghai 201800, China}
\affiliation{Graduate School of the Chinese Academy of Sciences,
Beijing 100080, China}
\author{ C. Zhong}
\affiliation{Shanghai Institute of Applied Physics, Chinese Academy of Sciences, P.O. Box 800-204, Shanghai
201800, China}
\author{ J. X. Zuo}
\affiliation{Shanghai Institute of Applied Physics, Chinese Academy of Sciences, P.O. Box 800-204, Shanghai
201800, China} \affiliation{Graduate School of the Chinese Academy of Sciences, Beijing 100080, China}

\date{ \today}

\begin{abstract}
The transverse momentum and pseudorapidity dependences of partonic
{`\it{Mach-like}'} shock wave have been studied by using a
multi-phase transport model (AMPT) with both partonic and hadronic
interactions. The splitting parameter $D$, i.e. half distance
between two splitting peaks on away side in di-hadron azimuthal
angular ($\Delta\phi$) correlations, slightly increases with the
transverse momentum of associated hadrons ($p^{assoc}_T$), which
is consistent with preliminary experimental trend, owing to
different interaction-lengths/numbers between wave partons and
medium in strong parton cascade. On the other hand, the splitting
parameter $D$ as a function of pseudorapidity of associated
hadrons ($\eta^{assoc}$), keeps flat in mid-pseudorapidity region
and rapidly drops in high-pseudorapidity region, which is as a
result of different violent degrees of jet-medium interactions in
the medium that has different energy densities in the longitudinal
direction. It is proposed that the research on the properties of
{`\it{Mach-like}'} correlation can shed light on the knowledge of
both partonic and hadronic interactions at RHIC.
\end{abstract}


\maketitle

An exotic matter predicted by Quantum ChromoDynamics
(QCD)~\cite{QCD} may be created in the early stage of central Au+Au
collisions at $\sqrt{s_{NN}}$ = 200 GeV at the Relativistic
Heavy-Ion Collider (RHIC) at the Brookhaven National
Laboratory~\cite{White-papers}. The sufficient experimental
evidences~\cite{ellipticflow,STARphi,jpsi,jet-ex} support that the
new matter is not an ideal gas but a strongly-interacting dense
partonic matter (named as sQGP) under extreme temperature and
energy density.

Jet has been proved to be a good probe to investigate the
characters of new matter in RHIC experiments~\cite{jet-ex}.
Calculations based on pQCD predicted that high energy partons
(jets) traversing a dense QCD medium lose energy through induced
gluon radiation~\cite{jet-quenching}, where the energy loss is
expected to depend strongly on the color charge density of the
created system and the traversed path length of the propagating
jets. The energy loss phenomenon (i.e. jet quenching) has been
observed by some experimental probes. In di-jet azimuthal angular
($\Delta\phi$) correlations, one back-to-back hard jet disappears
due to large energy loss in the dense medium~\cite{hard-hard-ex}.
According to the conservation law of energy, the lost energy
should be redistributed in the
medium~\cite{soft-soft-th1,soft-soft-th2,soft-soft-th3,soft-soft-th4},
which has been reconstructed by di-hadron azimuthal angular
correlations of charged particles in
experiments~\cite{soft-soft-ex}. Recently, an interesting
{`\it{Mach-like}'} structure (the splitting of the away side peak
in di-hadron $\Delta\phi$ correlation) has been observed in both
di-hadron and three-hadron azimuthal angular correlations in
central Au + Au collisions at $\sqrt{s_{NN}}$ = 200
GeV~\cite{sideward-peak1,sideward-peak2,sideward-peak3}, which is
attracting many theorists' attentions.

Recently considerable theoretical efforts have been put into the
topic~\cite{Casalderrey,Koch,large-angle,opaque-media-radiation,Ruppert}.
Among these researches, the transverse momentum ($p_{T}$) and
pseudorapidity ($\eta$) dependences of {`\it{Mach-like}'}
structure are two subjects of much concern. Cherenkov-like gluon
radiation model has been suggested to produce a conical
structure~\cite{Koch}. It gave the dispersion relations for
different choices of masses of two massive scalars coupled to a
massless scalar. The dispersion relations are space-like in low
momentum and approaches the light cone as momentum is increased.
With the dispersion relations, the cone angle with respect to the
jet direction becomes narrow rapidly with the increasing of the
momentum of the emitted particle. The medium-induced gluon
bremsstrahlung radiation~\cite{large-angle,opaque-media-radiation}
has been used to study the di-hadron correlations. In
Ref~\cite{large-angle}, the large angle medium-induced gluon
radiation, gave a picture of the radiative angle relative to the
jet axis becoming narrow with the increasing of radiative gluon
energy, but did not show a two-peak structure on away side. In
Ref~\cite{opaque-media-radiation}, a simple generalization of
Sudakov form factor to opaque media presented non-Gaussian shapes
of away-side in di-hadron azimuthal correlations and a centrality
dependence of the position of splitting peaks which can match
experimental data well. However, there are no discussions about
the $p_{T}$ dependence of non-Gaussian shape in the model. In a
shock-wave model proposed by Shuryak and his
collaborators~\cite{Casalderrey}, linearized hydrodynamic
equations were solved in a region far from the jet where the
perturbations and gradients are small. In general, the strength of
the diffusion mode relative to the sound mode is directly
proportional to the entropy produced by jet-medium interactions.
Only when there is no significant entropy produced by strong
jet-medium interactions, hydrodynamic fields can be induced and
peaks at the Mach angle in di-hadron azimuthal angular
correlations can be revealed. It is noticeable that the results
are very sensitive to the parameters chosen, especially need very
large values for the energy loss to produce the correlation
functions similar to the experimental results. The position of the
peak on away side, which corresponds to splitting parameter $D$
(the half distance between two peaks on away side in a di-hadron
azimuthal angular correlation), is related to the speed of sound
$c_{s}$. The value of $c_{s}$ used in the calculation is a
time-weighted average ($\bar{c_{s}}$ $\approx$ 0.33) consistent
with the expected value for RHIC. As a result, the positions of
the peaks at away side are located at $\Delta\phi$ $=$ $\pi$ $\pm$
arccos($\bar{c_{s}})$ $\approx$ 1.9, 4.3 (rad) and no $p_{T}$
dependence. On the other hand, a probable image of
{`\it{Mach-cone}'} shock wave was given as a function of
rapidity~\cite{renk-rapidity}, which implied that the splitting
parameter $D$ should become smaller with the increasing of the
rapidity (y), arising from a folding of condition possibility p(y)
and rapidity structure of {`\it{Mach-cone}'} correlation. In our
previous works, we have shown a partonic {`\it{Mach-like}'} shock
wave can be produced and evolved in strong parton cascade,
developed by hadronic rescattering, in two- and three-particle
azimuthal correlation
analysis~\cite{di-hadron,three-hadron,time-evolution}. In this
work, we report a study of $p_{T}$ and $\eta$ dependences of
partonic {`\it{Mach-like}'} shock waves by using A Multi-Phase
Transport model (AMPT)~\cite{AMPT}. The $p_{T}$ dependence of
splitting parameter ($D$) that $D$ increases with $p_{T}^{asso}$
slightly is attributed to different interaction-lengths/numbers
between wave partons and medium in strong parton cascade and
$\eta$ dependence of splitting parameter ($D$) that $D$ keeps flat
in mid-pseudorapidity and drops quickly in high-pseudorapidity is
due to different strengths of jet-medium interactions in the
medium which has different energy densities in the longitudinal
direction.

AMPT model~\cite{AMPT} is a hybrid model which consists of four
main components: the initial conditions, partonic interactions,
the conversion from partonic matter to hadronic matter and
hadronic rescattering. The initial conditions, which include the
spatial and momentum distributions of minijet partons and soft
string excitations, are obtained from the HIJING
model~\cite{HIJING}. Excitations of strings melt strings into
partons in the AMPT version with string melting
mechanism~\cite{SAMPT} (abbr. {`\it{the melting AMPT model}'}).
Scatterings among partons are modelled by Zhang's Parton Cascade
model (ZPC)~\cite{ZPC}, which at present only includes two-body
scatterings with cross sections obtained from the pQCD
calculations with screening mass. In the default version of AMPT
model~\cite{DAMPT}(abbr. {`\it{the default AMPT model}'}), minijet
partons are recombined with their parent strings when they stop
interactions, and the resulting strings are converted to hadrons
by using the Lund string fragmentation model~\cite{Lund}. In the
melting AMPT model, a simple quark coalescence model is used to
combine all partons into hadrons. Dynamics of the subsequent
hadronic matter is then described by A Relativistic Transport
(ART) model~\cite{ART}. Details of the AMPT model can be found in
a recent review~\cite{AMPT}. Previous studies have shown that the
partonic effect could not be neglected and the melting AMPT model
is much more appropriate than the default AMPT model when the
energy density is much higher than the critical density for the
predicted phase transition~\cite{AMPT,SAMPT,Jinhui}. In the
present work, the parton interaction cross section in the melting
AMPT model is assumed to be 10 mb.

Our analysis in this paper are based on di-hadron azimuthal
angular correlations between a high $p_{T}$ hadron (trigger
hadron) and low $p_{T}$ ones (associated hadrons). The analysis
method is similar to that the experimenters
did~\cite{soft-soft-ex,sideward-peak2}. The pairs of associated
and trigger particles in the same events were accumulated to
obtain $\Delta\phi = \phi_{assoc} - \phi_{trig}$ distributions.
Mixing-event technique was applied to construct background which
is expected mainly from elliptic
flow~\cite{soft-soft-ex,sideward-peak2}. In this method, we mixed
two events which have very close centralities into a new mixing
event, and obtained $\Delta\phi$ distribution which is regarded as
the corresponding background. A Zero Yield At Minimum (ZYAM)
assumption was adopted to subtract the background as did in
experimental analysis~\cite{sideward-peak2} (See
Ref.~\cite{di-hadron} for our detailed analysis techniques). In
this work, we took two sets of AMPT models (the {\it{default}} and
{\it{melting}} AMPT models) to simulate the central Au+Au
collisions (0-10$\%$) at $\sqrt{s_{NN}}$ = 200 GeV. The $p_{T}$
window cuts for trigger and associated particles are $3 <
p_{T}^{trig} < 4$ GeV/$c$ and $0 < p_{T}^{assoc} < 3$ GeV/$c$
within pseudorapidity window $|\eta^{trig,assoc}| < 1$ for $p_{T}$
dependence analysis, and $2.5 < p_{T}^{trig} < 6$ GeV/$c$ and $1 <
p_{T}^{assoc} < 2.5$ GeV/$c$ within pseudorapidity window
$|\eta^{trig}| < 1$ for $\eta^{assoc}$ dependence analysis.
%
\begin{figure}[htbp]
\resizebox{8.6cm}{!}{\includegraphics{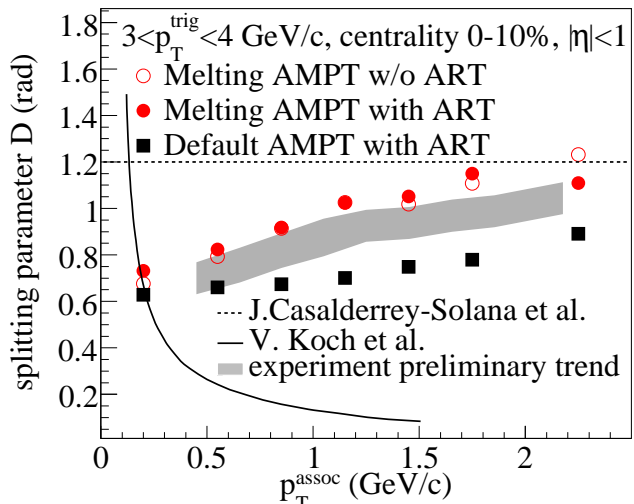}} \vspace{-0.8cm}
\caption{(Color online) The splitting parameter $D$ as a function
of $p^{assoc}_T$ for trigger hadrons of $3 < p_{T}^{trig} < 4
$GeV/$c$ from AMPT simulations for Au + Au collisions (0-10\%) at
$\sqrt{s_{NN}}$ = 200 GeV. Full circles: the melting AMPT model
with hadronic rescattering (i.e. ART process); open circles: the melting AMPT model
without hadronic rescattering; full squares: the default AMPT
model with hadronic rescattering; band: preliminary experimental
trend~\cite{Mark-STAR}. } \label{D_pT_figure} \vspace{-0.5cm}
\end{figure}
\begin{figure}[htbp]
\resizebox{8.6cm}{!}{\includegraphics{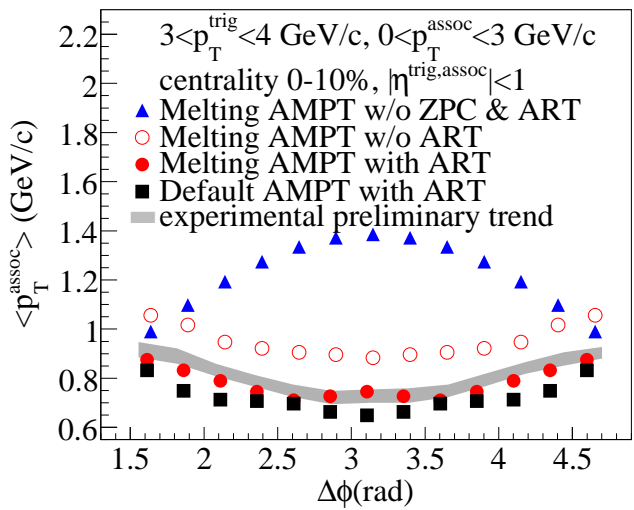}} \vspace{-0.8cm}
\caption{(Color online) AMPT model calculations of $<p^{assoc}_T>$
versus $\Delta\phi$ for away side in di-hadron $\Delta\phi$
correlations with $p_T$ window cut ( $3 < p^{trig}_T <4$ GeV/$c$
and $0 < p_{T}^{assoc} < 3$ GeV/$c$) in Au + Au collisions
(0-10\%) at $\sqrt{s_{NN}}$ = 200 GeV. Full circles: the melting
AMPT model with hadronic rescattering ; open circles: the melting
AMPT model without hadronic rescattering; full triangles: the
melting AMPT model without parton cascade (i.e. ZPC process) and hadronic rescattering; full squares: the default AMPT model with hadronic
rescattering; band: preliminary experimental
trend~\cite{Mark-STAR}. } \label{meanpT_figure} \vspace{-0.5cm}
\end{figure}

\vspace{0.5cm}
 Figure~\ref{D_pT_figure} shows $p_{T}^{assoc}$
dependences of splitting parameter $D$ in Au + Au collisions
(0-10\%) at $\sqrt{s_{NN}}$ = 200 GeV  in the melting and default
AMPT model with/without hadronic rescattering (i.e. ART process). Here the trigger
and associated particles are with $3 < p_{T}^{trig} < 4$ GeV/$c$
and $0 < p_{T}^{assoc} < 3$ GeV/$c$ respectively, both within
pseudorapidity window $|\eta^{trig,assoc}| < 1$. There is no
significant difference of splitting parameter $D$ between the
cases with and without hadronic rescattering in the melting AMPT
model, but $D$ in the default AMPT model is smaller than that in
the melting AMPT model. It is consistent with our previous
results~\cite{di-hadron} in which hadronic rescattering alone
gives much smaller splitting amplitudes than that in the melting
AMPT model which shows good $D$ values matching experimental data.
Here the melting AMPT model gives much better splitting parameters
$D$ which are consistent with the preliminary experimental
trend~\cite{Mark-STAR}, but different from the calculations of
Casalderrey-Solana et al.~\cite{Casalderrey} and Koch et
al.~\cite{Koch}. It is noticeable that the $p_{T}^{assoc}$
dependence of splitting parameter $D$ is sensitive to parton
cascade (Note that it was found that more long-lived partonic
phase and bigger parton interaction cross section can give more
bigger D values in our previous work~\cite{time-evolution}.), but
not to hadronic rescattering in the melting AMPT model. Therefore,
the relation of D vs. $p_{T}^{assoc}$ can be a good probe that not
only distinguishes different physical scenarios but also sheds
light on the properties of partonic interactions in the early
stage of the relativistic heavy ion collisions.

\begin{figure}[htbp]
\resizebox{8.6cm}{!}{\includegraphics{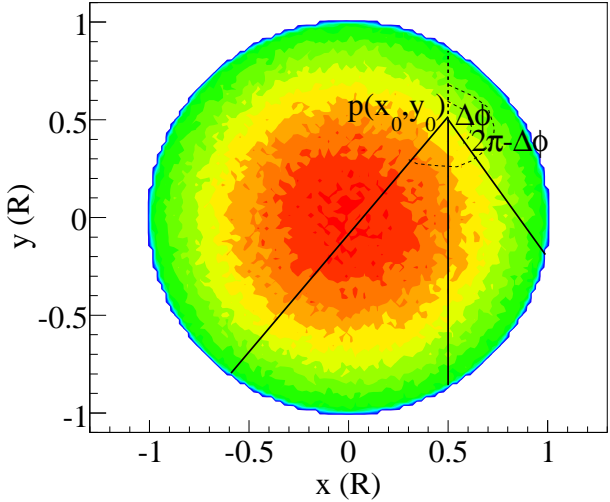}} \vspace{-0.8cm}
\caption{(Color online) The initial distribution of parton number
density in mid-rapidity for Au + Au collisions (0-10\%) at
$\sqrt{s_{NN}}$ = 200 GeV in the melting AMPT model, which has
been projected into x-y plane, where p($x_{0}$,$y_{0}$) is the
generation point of Mach-like shock wave and R is the scale size
of the system.} \label{circle-density} \vspace{-0.5cm}
\end{figure}

\begin{figure}[htbp]
\resizebox{8.6cm}{!}{\includegraphics{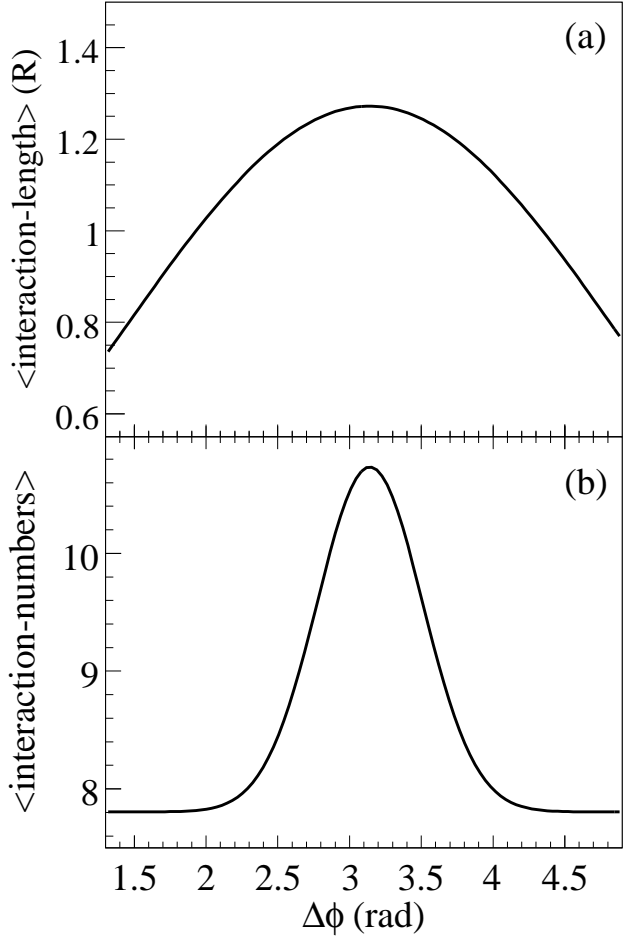}}\vspace{-0.5cm}
\caption{(a): The average interaction-lengths of a wave parton
undergoing the medium in different $\Delta\phi$; (b): The average
interaction-numbers of a wave parton undergoing the medium in
different $\Delta\phi$ for Au + Au collisions (0-10\%) at
$\sqrt{s_{NN}}$ = 200 GeV in the melting AMPT model.}
\label{length-numbers} \vspace{-0.5cm}
\end{figure}

Figure~\ref{meanpT_figure} presents the relation of average
transverse momentum of associated particles ($<p_{T}^{assoc}>$
$\equiv$ $\frac{\Sigma_{i}N^{i}_{\ assoc}p_{T,assoc}^{i}}
{\Sigma_{i}N^{i}_{\ assoc}}$ ) versus $\Delta\phi$ for away side.
There is a distinct dip structure at $\Delta\phi$ = $\pi$ for all
model conditions except the melting AMPT model without parton
cascade and hadronic rescattering, which indicates that away jets
are depleted by strong parton cascade in the melting AMPT model
(or by hadronic rescattering in the default model) and its lost
energy is expected to excite the formation of partonic (or
hadronic) {`\it{Mach-like}'} shock waves~\cite{time-evolution}. It
is apparent that hadronic rescattering can soften the associated
hadrons by comparing the results between with and without hadronic
rescattering in the melting AMPT model. The depletion structure
implies that harder associated hadrons prefer larger angles with
respect to the away-jet direction, which results in the increasing
trend of splitting parameter vs $p_{T}^{assoc}$ as shown in
Figure~\ref{D_pT_figure}. In order to explain the dip structure,
we make a simple MC simulation. Figure~\ref{circle-density} gives
the initial parton number density which is projected into $x-y$
plane in the melting AMPT model. Once shock wave is generated at
some points $p(x_{0},y_{0})$, the wave-front direction would be
symmetrical with respect to the away-jet direction (at
$\Delta\phi=\pi$ rad). The average interaction-lengths/numbers of
wave-front partons interacting with the medium are with maximums
at $\Delta\phi=\pi$ rad (i.e. in the away-jet direction), as shown
in panel (a) and (b) of Figure~\ref{length-numbers}, which is the
cause of the dip structure of $<p_{T}^{assoc}>$ vs $\Delta\phi$
for away side. Though $<p_{T}^{assoc}>$ vs. $\Delta\phi$ in the
final state is a result of both parton cascade and hadronic
rescattering in the melting AMPT model, it is interesting that a
combination research between $<p_{T}^{assoc}>$ vs $\Delta\phi$ and
D vs. $p_{T}^{assoc}$ may provide a potential tool to extract the
respective information from two different interaction levels (i.e.
partonic and hadronic stages).

As we have already shown that the origin of partonic
{`\it{Mach-like}'} shock waves could be attributed to the big
partonic interaction cross section which can couple partons
together to show a hydro-like collective behavior in our previous
papers~\cite{three-hadron,time-evolution}, and our used model is a
MC dynamical partonic transport model which can harmoniously work
in the regions close to and far from jet, whereas linear
hydrodynamical approximation is only applicable in the region far
from jet as discussed above. Therefore, it is not strange that
such an increasing dependence, which is different from the
predicted by a hydrodynamic calculation~\cite{Casalderrey}, is
presented in the present model.

\begin{figure}[htbp]
\resizebox{8.6cm}{!}{\includegraphics{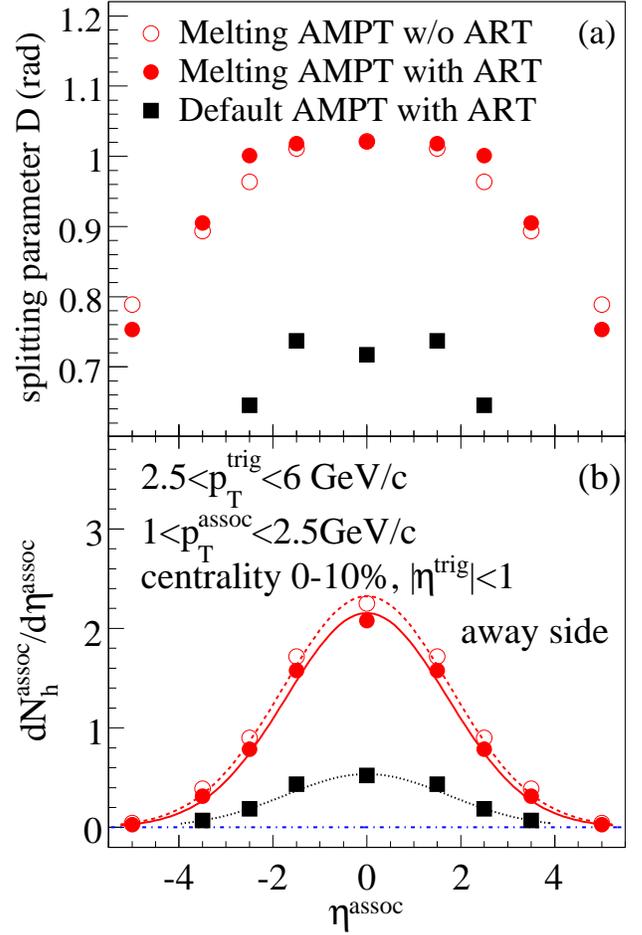}}
\vspace{-0.8cm} \caption{(Color online) The AMPT model
calculations in Au + Au collisions (0-10\%) at $\sqrt{s_{NN}}$ =
200 GeV. (model cuts: $2.5 <p_{T}^{trig}< 6$ GeV/$c$, $1
<p_{T}^{assoc}< 2.5$ GeV/$c$, and $|\eta^{trig}|<1$). (a): the
pseudorapidity dependences of splitting parameter $D$; (b): the
pseudorapidity distributions of associated hadron yields for away
side. Full circles: the melting AMPT model with hadronic rescattering (i.e. ART process); open circles: the melting AMPT model without
hadronic rescattering; full square: the default AMPT model with
hadronic rescattering; lines: the corresponding Gaussian fitting
functions. } \label{D_N_eta_figure} \vspace{-0.5cm}
\end{figure}
%
\begin{figure}[htbp]
\resizebox{8.6cm}{!}{\includegraphics{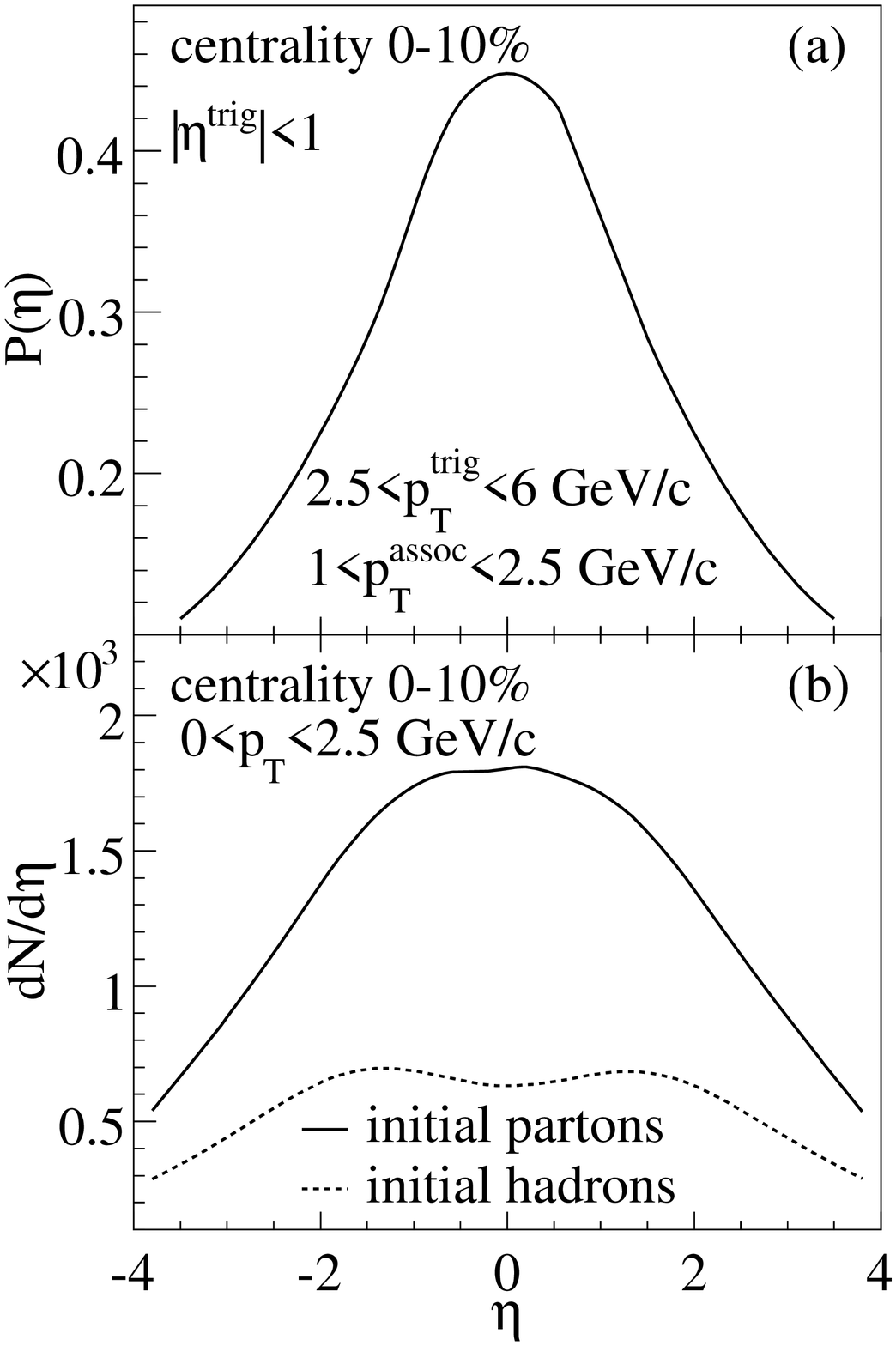}}
\vspace{-0.8cm}
 \caption{AMPT model calculations in Au + Au collisions (0-10\%) at $\sqrt{s_{NN}}$ = 200 GeV.
(a): conditional probability $P(\eta)$ to find the away side
associated parton $(1 < p_{T}^{assoc} < 2.5 GeV/c)$ at
pseudorapidity $\eta$ if the trigger parton $(2.5 < p_{T}^{trig} <
6 GeV/c)$ is found at $|\eta^{trig}|<1$ in the initial state of
melting AMPT model; (b): the pseudorapidity distribution of
initial partons (hadrons) $(0 < p_{T} < 2.5 GeV/c)$ before parton
cascade (hadronic rescattering). solid: $\eta$ distribution of
initial partons in the melting AMPT model; dash: $\eta$
distribution of initial hadrons in the default AMPT model. }
\label{dN_dy_p_eta_figure} \vspace{-0.5cm}
\end{figure}

In the above studies on $p_{T}$ dependences of $D$, both the
trigger and associated hadron both are located at
mid-pseudorapidity ($|\eta^{trig,assoc}| < 1$). When one scans
the whole pseudorapidity region to look for all hadrons which are
associated with a mid-pseudorapidity trigger hadron, how about the
associated hadrons in other pseudorapidity regions? Are they waked
by an away-jet? Figure~\ref{D_N_eta_figure} (a) shows the
dependences of splitting parameter $D$ on the pseudorapidity of
associated hadrons ($\eta^{assoc}$), for a given trigger hadron
within $|\eta^{trig}|<1$. The splitting parameter $D$ is almost
unchanged in a wide mid-pseudorapidity region, but goes down
quikly at high absolute value of pseudorapidity. For the splitting
parameter $D$ in the default AMPT model with hadronic
rescattering, it is lower than that in the melting AMPT model and
not comparable with experimental results, which further
demonstrates that strong parton cascade process is necessary for
reproducing reasonable splitting parameters.
Figure~\ref{D_N_eta_figure} (b) gives pseudorapidity distributions
of associated hadron yields for away side with a trigger hadron in
mid-pseudorapidity. The associated yield on away side in the
melting AMPT model is higher than that in the default AMPT model,
but there is no distinct difference between the two cases in the
melting AMPT model with/without hadronic rescattering. It is
interesting that the associated away-side yields in the melting
AMPT model can be extended to high pseudorapidity region ($|\eta|$
$>$ 4), but those in the default AMPT model are confined within
the region $|\eta|$ $<$ 4, which indicates that the medium at
high pseudorapidity can be waked by an away-side jet due to strong
parton interactions.

In panel (a) of Figure~\ref{dN_dy_p_eta_figure}, it shows the
conditional probability $P(\eta)$ to find the away side associated
parton $(1 < p_{T}^{assoc} < 2.5 GeV/c)$ at some pseudorapidity
$\eta$, if a trigger parton $(2.5 < p_{T}^{trig} < 6 GeV/c)$ is
found at $|\eta^{trig}|<1$ in the initial state before parton
cascade, which is consistent with the LO pQCD
calculations~\cite{renk-rapidity}. It means that the away-jet is
with a certain possibility to be in non-mid-pseudorapidity. At the
same time, the pseudorapidity ($\eta$) distribution of initial
partons (hadrons) $(0 < p_{T} < 2.5 GeV/c)$ before parton cascade
(hadronic rescattering) is also presented in panel (b) of
Figure~\ref{dN_dy_p_eta_figure}. Since $\eta$ distribution of
particle number is proportional to the $\eta$ distribution of
energy density, the interactions between away side jets and
associated particles should be more violent in mid-pseudorapidity
region than in high $|\eta|$ region. In our previous
work~\cite{time-evolution}, the formation of partonic
{`\it{Mach-like}'} shock waves stems from strong parton cascade,
therefore the $\eta^{assoc}$ dependence of splitting parameter is
expected to result from different violent degrees of jet-medium
interactions in the medium which has different energy densities in
the longitudinal direction. From Figure~\ref{D_N_eta_figure} (a) and
(b), the $\eta^{assoc}$ dependence of the splitting prameter can
identify the partonic or hadronic origin of the Mach-like
behavior, and provide the information about minijet production
mechanism and the formed matter density distribution in the
longitudinal direction.

In conclusion, the properties of partonic {`\it{Mach-like}'} shock
waves, including $p_{T}^{assoc}$ and $\eta^{assoc}$ dependences,
have been investigated in the framework of a hybrid dynamical
transport model which consists of two dynamical processes, namely
parton cascade and hadronic rescaterring. It was found that the
splitting parameter $D$ increases slightly with the transverse
momentum of associated hadrons, which results from different
interaction-lengths/numbers between wave partons and medium in
strong parton cascade. On the other hand, the pseudorapidity
dependence of splitting parameter $D$ induced by a
mid-pseudorapidity trigger hadron keeps almost a constant in
mid-pseudorapidity region but drops rapidly in higher
pseudorapidity region, which is on account of different strength
of jet-medium interactions in the medium which has different
energy densities in the longitudinal direction. Therefore, the
correlative researches on the properties ($p_{T}$ and $\eta$
dependences) of {`\it{Mach-like}'} correlation can provide a
potential tool to explore the characters of both partonic and
hadronic interactions in the formed matter at RHIC.

This work was supported in part by the Shanghai Development
Foundation for Science and Technology under Grant No. 05XD14021,
the National Natural Science Foundation of China under Grant No.
10610285 and 29010702 and the Knowledge Innovation Project of the Chinese
Academy of Sciences under Grant No. KJCX2-YW-A14 and and
KJXC3-SYW-N2.  And we thank Information Center of Shanghai
Institute of Applied Physics of Chinese Academy of Sciences for
using PC-farm.

\end{document}